\documentclass[journal]{IEEEtran}

\usepackage[utf8x]{inputenc}
\usepackage{bm}
\usepackage{amsfonts}
\usepackage{amsmath}
\usepackage{txfonts}
\usepackage{amssymb}
\usepackage[compress]{cite}
\usepackage{graphicx}
\usepackage{algorithm}
\usepackage{algorithmic}
\usepackage{float}
\usepackage{subfigure}
\usepackage{array}
\usepackage{booktabs}
\usepackage{bookmark}
\usepackage{multirow}
\usepackage{epsfig}
\usepackage{url}
\usepackage{enumerate}
\usepackage{color}
\usepackage[table]{xcolor}
\newcommand{\vect}[1]{\boldsymbol{\mathbf{#1}}}

\makeatletter

\def\ps@IEEEtitlepagestyle{%
  \def\@oddfoot{\mycopyrightnotice}%
}

\def\mycopyrightnotice{%
}
\makeatother

\makeatother

\ifCLASSINFOpdf

\else

\fi

\usepackage{stfloats}

\begin{document}
\bstctlcite{IEEEexample:BSTcontrol}
%
\title{When Provably Secure Steganography Meets Generative Models}

\author{Kejiang Chen, Hang Zhou, Hanqing Zhao, Dongdong Chen, Weiming Zhang, Nenghai Yu
\thanks{
This work was supported in part by the Natural Science Foundation of China under Grant U1636201, 61572452.}
\thanks{All the authors are with CAS Key Laboratory of Electro-magnetic Space Information, University of Science and Technology of China, Hefei 230026, China. }
\thanks{Corresponding author: Weiming Zhang (Email: zhangwm@ustc.edu.cn)}
}

\maketitle
\begin{abstract}
Steganography is the art and science of hiding secret messages in public communication so that the presence of the secret messages cannot be detected. There are two provably secure steganographic frameworks, one is black-box sampling based and the other is compression based. The former requires a perfect sampler which yields data following the same distribution, and the latter needs explicit distributions of generative objects. However, these two conditions are too strict even unrealistic in the traditional data environment, because it is hard to model the explicit distribution of natural image. With the development of deep learning, generative models bring new vitality to provably secure steganography, which can serve as the black-box sampler or provide the explicit distribution of generative media. Motivated by this, this paper proposes two types of provably secure stegosystems with generative models. Specifically, we first design block-box sampling based provably secure stegosystem for broad generative models without explicit distribution, such as GAN, VAE,  and flow-based generative models, where the generative network can serve as the perfect sampler. 
For compression based stegosystem, we leverage the generative models with explicit distribution such as autoregressive models instead, where the adaptive arithmetic coding plays the role of the perfect compressor, decompressing the encrypted message bits into generative media, and the receiver can compress the generative media into the encrypted message bits.
To show the effectiveness of our method, we take DFC-VAE, Glow, WaveNet as instances of generative models and demonstrate the perfectly secure performance of these stegosystems with the state-of-the-art steganalysis methods. 

\end{abstract}


\IEEEpeerreviewmaketitle

\section{Introduction}

\IEEEPARstart{S}{teganography} is the art and science of communicating in such a way that the presence of a message cannot be detected. It has been be applied in various applications, such as anonymous communication and covert communication, by hiding message into media, such as text\cite{yang2018rnn}, audio\cite{chen2019derivative}, image\cite{su2018new,yeung2019secure}, video\cite{wang2018maintaining}, etc. Cachin\cite{cachin1998information} firstly formalized an information-theoretic model for steganography in 1998, where a relative entropy function is employed as a basic measure of steganographic security for the concealment system. The security of a steganographic system can be quantified in terms of the relative entropy $D\left(P_{\text{c}} \parallel P_{\text{s}} \right)$ between the distributions of \emph{cover} $P_{\text{c}}$ and \emph{stego} $P_{\text{s}}$, which yields bound on the detection capability of any adversary. From another perspective, Hopper \emph{et al.}\cite{hopper2002provably} formalized perfectly secure system based on computational-complexity, which is immigrated from cryptography. The perfect security is defined as a polynomial-time distinguisher that cannot distinguish the \emph{cover} and \emph{stego}.

With the definition of steganographic security, provably secure steganography has developed a lot in the past two decades and can be divided into two frameworks: compression based and black-box sampling based. Anderson \emph{et al.}\cite{anderson1998limits} observed that \emph{cover} can be compressed to generate a secret message, and for embedding message, we can decompress it into \emph{stego}. Le \emph{et al.}\cite{van2003efficient} constructed a provably secure steganography called $\mathcal{P}$-code based on arithmetic coding and it assumes that the sender and receiver know the distribution of \emph{cover} exactly. Sallee \cite{sallee2003model} designed a compression based stegosystem for JPEG images that assumes the Alternating Current (AC) coefficients in JPEG images following Generalized Cauthy distribution, and the receiver can estimate the distribution as well. 

As for black-box sampling based stegosystem, Cachin\cite{cachin1998information} proposed rejection-sampling to generate \emph{stego} that looks like \emph{cover}. In detail, the stegosystem samples from the \emph{cover} distribution until it selects a document whose hash value equals to message XOR $k$, where $k$ is a session key both parties derive from the shared secret key. Hopper\cite{hopper2002provably} improved Cachin's method and generalized it to be applicable to any distribution, which assumes it has sufficient entropy and can be sampled perfectly based on prior history. Von Ahn \emph{et al.}\cite{von2004public} created public-key provably secure stegosystem and chosen-stegotext attacks (SS-CSA). Lysyanskaya \emph{et al.}\cite{lysyanskaya2006provably} analyzed the problem of imperfect sample by weakening assumption, where the \emph{cover} distribution is modeled as a stateful Markov process. Zhu \emph{et al.}\cite{zhu2012efficient} provided a more general construction of secure steganography with one-way permutation and unbiased sampler.

Nevertheless, these two types of steosystems are both not effective and not feasible in the realistic traditional data environment. For compression based systems, they need to know the exact distribution of \emph{cover}, which is too strict. Because the complexity and dimensionality of \emph{covers} formed by digital media objects, such as natural audios, images and videos, prevent us from modeling a complete distribution $P_{\text{c}}$ of \emph{cover}. As for black-box sampling based system, the difficulty is that the perfect sampler is hard to obtain, and the capacity of the existing schemes is rather low. 

With the development of deep learning, generative models has obtained significant success in recent years and bring new vitality to provably secure steganography. By approaching the distribution of training datasets, they can generate diverse data samples in the test stage. Prominent models include variational auto-encoders (VAE)\cite{kingma2013auto,rezende2014stochastic}, which maximize a variational lower bound on the data log likelihood, generative adversarial networks (GAN)\cite{goodfellow2014generative}, which employs an adversarial framework to train a generative model that mimics the true transition model, and auto-regressive models\cite{germain2015made,van2016conditional,oord2016wavenet} and normalizing flows\cite{kingma2018glow} which train with maximum likelihood (ML), avoiding approximations by choosing a tractable parameterization of probability density. 

For VAE, GAN, and flow-based generative models, they can generate vivid objects from latent variables, which follow a prior distribution, e.g. normal distribution. For auto-regressive models, they can give explicit distribution of generative objects instead.  These properties solve the difficulties in the traditional data environment, showing a new direction of designing provably secure steganography. Note that, though generative models have been utilized for steganography in some recent works\cite{hayes2017generating,hu2018novel,zhu2018hidden}, most of them do not focus on provably secure steganography.

In this paper, we design provably secure stegosystems on generative models, which owns the advantages of efficiency, practicality and high capacity. As for VAE, GAN and flow-based models, since they cannot provide implicit distribution of generative media but can be seen as perfect black-box samplers, so they can be leveraged to design black-box-sample based stegosystems. In detail, the encrypted message following uniform distribution is mapped into the latent vector which follows normal distribution first, and then the latent vector is fed to the generative component, yielding the generative data. To extract messages, another neural network whose structure is similar to the discriminator in GAN or encoder in VAE should be trained. For flowed-based generative models, thanks to their inherent reversible structures, they can directly recover message from generated data. If the parameters of generative models are kept unchanged, feeding the latent vector following the specified distribution (e.g., normal distribution), the distribution of the generated data will be same thus guarantee the perfect security. In the experiment part, DFC-VAE and Glow are taken as the realizations for building the stegosystem, and steganalytic analysis is further used to verify the final secure performance. 

When it comes to auto-regressive models, since they can express tractable distribution of generative data, which meets the requirement of compression based stegosystem. Therefore, compression based stegosystem are adopted for these models. In this paper, we use one of most representative generative model (i.e., WaveNet) as an example to construct the system. This model is originally designed to synthesize high quality audios, and the semantic of generated audios is very robust. Specifically, to build the system, we combine arithmetic coding into the generation process. In the sender end, it decompresses the encrypted message into audio samples following the given distribution predicted by WaveNet. Then the receiver can reproduce the same probability distribution as that in the sender end with the same generative model. Finally, by feeding the distribution and the \emph{stego}, the message can be extracted by the encoder of arithmetic coding. In the following part, we will provide the theoretical analysis of the stegosystem using arithmetic coding to prove the perfect security.

To sum up, the main contributions of this work are three-fold:
\begin{itemize}

	\item We introduce VAEs, GANs and flow-based generative models for black-box sampling based provably secure steosystems. 

	\item Based on auto-regressive generative model WaveNet, we design the compression based provably secure stegosystem. Benefit from the robustness of the semantics of audio, the receiver can resume the generative process without additional information. Theoretical analysis is also given to prove the perfect security.
	
	\item Extensive experiments demonstrate the perfect security proposed stegosystems by defending against the state-of-the-art steganalytic methods. Most of the proposed stegosystems show perfect secure performance.
	
\end{itemize}
The rest of this paper is organized as follows. We review the related work in Section II, and the proposed provably secure stegosystem based on different kinds of generative models are elaborated in Section III and Section IV respectively. The experiments are presented in Section V. Conclusion and future work are given in Section VI.

\section{Related Work}

\subsection{The Prisoners' Problem}
In order to improve the readability of this paper, the prisoners' problem\cite{simmons1984prisoners} formulated by Simmons is first introduced, which is a simple simulation on the background of steganography - Alice and Bob are imprisoned in separate cells and want to hatch an escape plan. They are allowed to communicate but their communication is monitored by warden Eve. If Eve finds out that the prisoners are secretly exchanging the messages, she will cut the communication channel and throw them into solitary confinement.
\begin{figure}[t]
	\centering
	\includegraphics[width=3.5in]{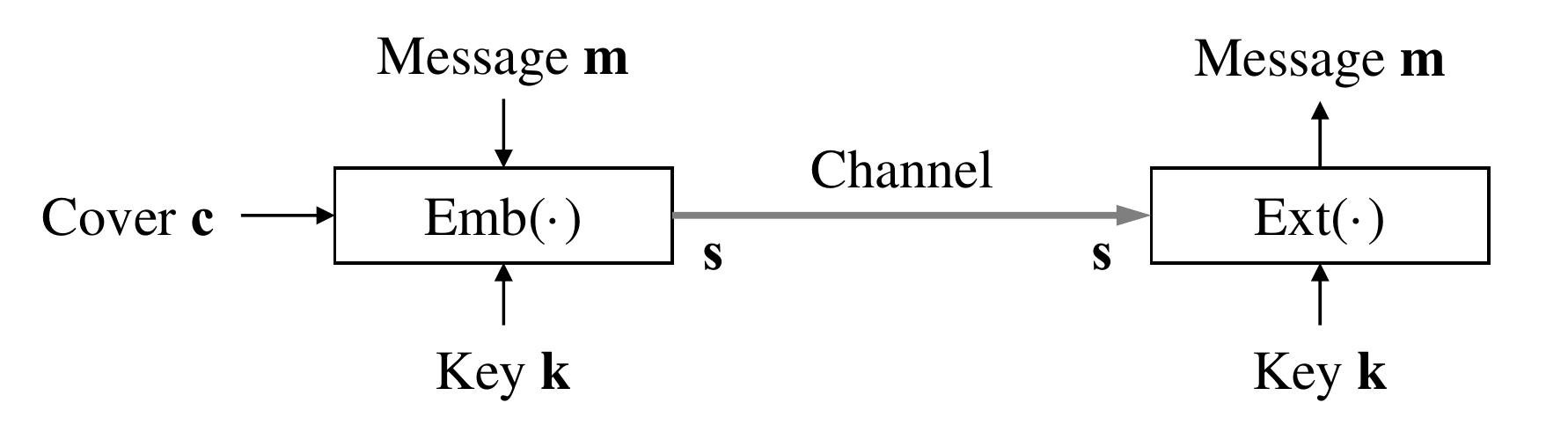}
	\caption{A diagram of steganographic communication.}
	\label{diagram}
\end{figure}

\subsection{The Diagram of Steganography}
According to the prisoners' problem, the diagram of the steganography is depicted in Fig. \ref{diagram}. A steganographic scheme can be regarded as a pair of embedding and extraction functions Emb() and Ext() for Alice and Bob, respectively\cite{fridrich2009steganography}.
\begin{equation}
\text{Emb}(\vect{c},\vect{k},\vect{m}) = \vect{s},
\end{equation}
\begin{equation}
\text{Ext}\left(\vect{s}\right)=\vect{m},
\end{equation}
where $\vect{c}, \vect{k},\vect{m},\vect{s}$ are \emph{cover} object, secret key, message and \emph{stego} object, respectively.
Eve judges the object $\vect{s}$ is innocent or not by all the possible knowledge except secret key according to Kerckhoffs' principle\cite{kerckhoffs1883kerckhoffs}.


\subsection{The Theoretical Definition of Steganographic Security}
The theoretical definition of steganographic security is given by Cathin\cite{cachin1998information} with the information theory. Assuming the \emph{cover} is drawn from $\mathcal{C}$ with probability distribution $P_\text{c}$ and the steganographic method will generate \emph{stego} object which has the distribution $P_\text{s}$. 
The distance of two distributions can be measured using relative entropy:
\begin{equation}
D\left( P_{\text{c}}\parallel P_{\text{s}} \right) = \sum_{\vect{x} \in \mathcal{C}}P_{\text{c}}\left(x\right) \log \frac{P_{\text{c}}\left(x\right)}{P_{\text{s}}\left(x\right)}.
\end{equation}
When $D\left( P_{\text{c}}\parallel P_{\text{s}} \right) = 0$, the stegosystem is perfectly secure. 

\subsection{Existing Provably Secure Stegosystem}
The black-box sampling based provably secure stegosystem can be briefly described in Algorithm \ref{blackbox}.
Given a mapping function $f_{k}(\cdot) : \mathcal{K} \times \mathcal{C} \rightarrow \mathcal{R}$ with the secret key $k$ and $\mathcal{R}=\{0,1\}^{e}$, the message embedding in the stegosystem is based on rejection sampling algorithm $\emph{Sample}_{f}^{\mathcal{C}}(k, b)$. The algorithm can sample \emph{cover} following the given distribution $\mathcal{C}$ by oracle $\mathcal{O}^{\mathcal{C}}$, such that an $e$-bit symbol $b$ will be embedded in it. The algorithm randomly chooses a sample $s$ until $f_{k}(s)=b$, where $\leftarrow_R$ denotes the random choosing of oracle $\mathcal{O}$. Nevertheless, the perfect samplers for multimedia are hard to obtain in the traditional data environment, and the capacity of existing schemes, such as adopting document, network protocol\cite{hopper2009provably} as the carrier, is rather low, e.g. one document carries 1 bit message. In addition, there seems no black-box sampling based perfectly secure steganography for multimedia.
\begin{algorithm}
	\caption{$\emph{Sample}^{\mathcal{C}}_{f}(k, b)$}
	\label{blackbox}
	\begin{algorithmic}[1]
		\REQUIRE  $\text{a key}\ k, \text{ a value } b \in\{0,1\}^{e}$
		\REPEAT
		\STATE  $s \leftarrow_{R} \mathcal{O}^{\mathcal{C}}$
		\UNTIL $f_{k}(s)=b$
		\STATE \text{Return} s
	\end{algorithmic}
\end{algorithm}

\begin{figure*}[t]
	\centering
	\includegraphics[width=5in]{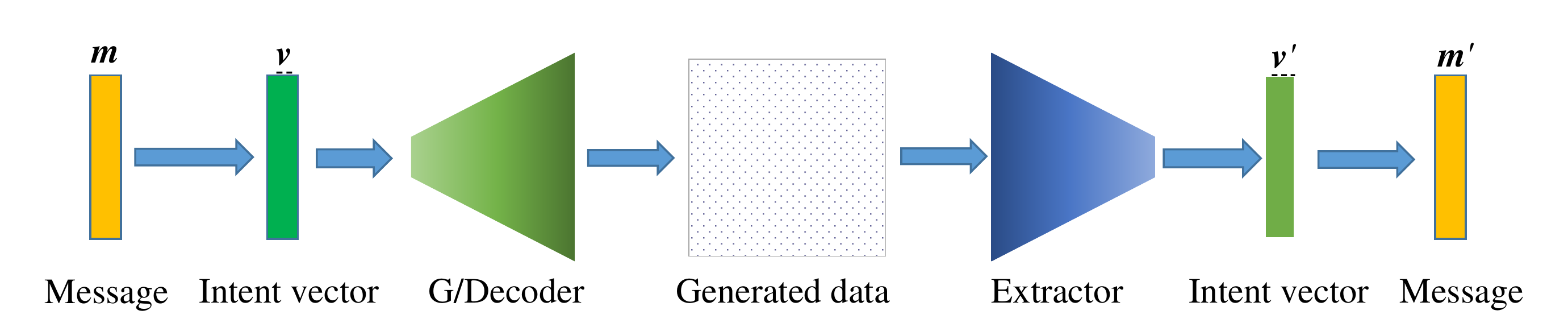}
	\caption{The pipeline of black-box sample based provably secure steganography using VAE or GAN.}
	\label{blackframework}
\end{figure*}{}

Based on arithmetic coding, Le \emph{et al.}\cite{van2003efficient} proposed $\mathcal{P}$-code for provably secure steganography, which requires both sides know the distribution of \emph{cover} exactly. Sallee \cite{sallee2003model} designed the compression-based stegosystem for JPEG images by assuming Alternating Current (AC) coefficients follow Generalized Cathin distribution. However, the complexity and dimensionality of covers formed by digital media objects, such as natural audios, images and videos, will prevent us from determining a complete distribution $P_{\text{c}}$ of cover, which implies Sallee's method cannot achieve perfectly secure. 

In summary, compression-based schemes need to know the exact distribution of cover, which is too difficult to capture the distribution of digital media objects. As for black-box-sample based system, the perfect sampler are hard to obtain, and the capacity of the existing schemes is rather low. To this end, we introduce the generative models into provably secure steganography which can serve as the perfect sampler, or supply the explicit probability distribution of \emph{cover} object, so that the provably secure stegosystem can be practical and effective. 

\subsection{Generative Models}\label{fram}
\begin{figure}[h]
	\centering
	\includegraphics[width=1\columnwidth]{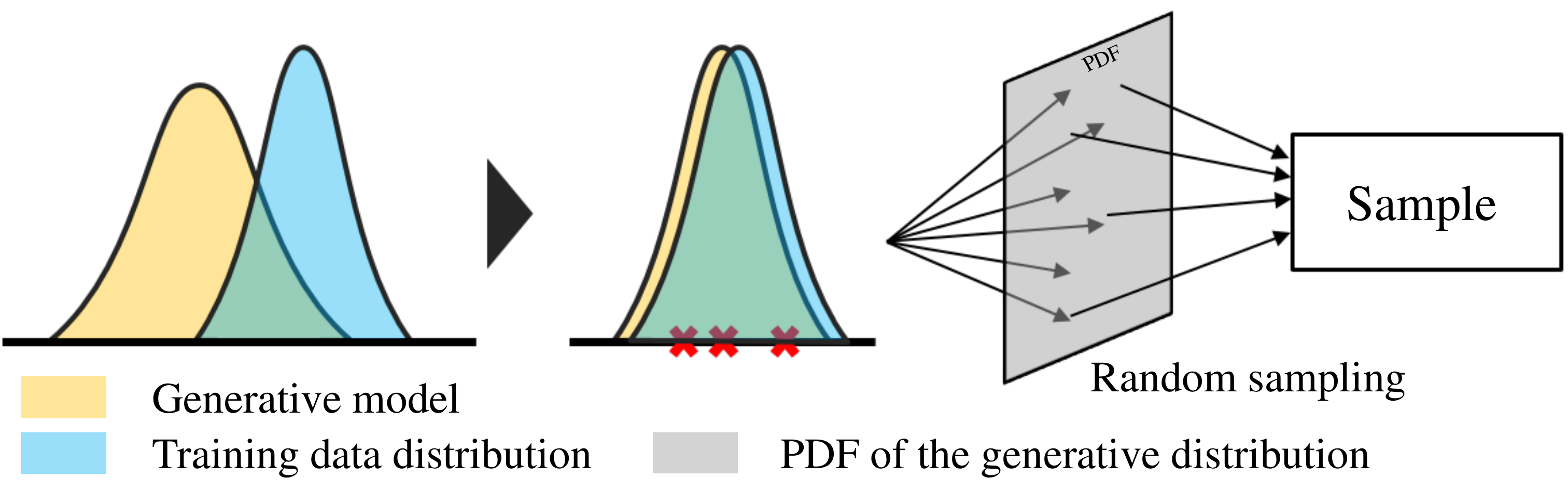}
	\caption{The pipeline of sample generation using generative model.}
	\label{generative}
\end{figure}

Generative model describes how media are generated, in terms of a probabilistic model. The generation process can be seen as random sampling from the probability distribution learned in the stage of training, as shown in Fig. \ref{generative}. The generative models can be divided into two categories, implicit density probability distribution and explicit density probability distribution. VAEs, GANs and flow-based models belong to the first category, and auto-regressive models attribute to the second category. The former meets the requirement of the black-box sampling based stegosystem, while the latter is able to be adopted to develop compression based stegosystem. In the subsequent sections, we will design provably secure stegosystem on different generative models.

\section{Provably Secure Steganography on Generative Model Without Explicit Probability Distribution}\label{s3}
VAEs, GANs and flow-based models do not own explicit probability distribution, however, the generative components in their models can serve as the perfect black-box sampler, which generate media with the same distribution. The black-box sampling based provably secure stegosystem can be built on them. Details will be expanded below.

\subsection{Provably Secure Stegosystem Using VAE / GAN}\label{s31}
VAE includes \emph{encoder} and \emph{decoder}. To generate a sample from the model, the VAE first draws a sample $\vect{z}$ from the prior distribution $p_{\theta}(\mathbf{z})$. The sample $\vect{z}$ is used as input to a differentiable generator network $g(\mathbf{z})$. Finally, $\vect{x}$ is sampled from a distribution $p_{\theta}(\mathbf{x} | g(\mathbf{z}))=p_{\theta}(\mathbf{x} | \mathbf{z})$.  During the training, the approximate inference network, \emph{i.e.}, encoder network $q_{\phi}(\mathbf{z} | \mathbf{x})$ is  used to obtain $\vect{z}$ and $p_{\theta}(\mathbf{x} | \mathbf{z})$ is viewed as a decoder network.

GAN is one of the most popular generative deep models. The core of a GAN is to play a min-max game between a discriminator D and a generator G, \emph{i.e.}, adversarial training. The discriminator D tries to differentiate if a sample is from real-world or generated by the generator while the generator G tries to generate samples that can fool the discriminator. The generator takes a noise $\mathbf{z}$ sampled from a prior distribution $p(\mathbf{z})$ as input, and maps the noise to the data space as $G_{\theta_{g}}\left(\mathbf{z}\right)$. Typical choices of the prior $p(\mathbf{z})$ can be uniform distribution or Gaussian distribution. $D_{\theta_{d}}\left(\mathbf{x}\right)$ is the classifying network that outputs a single scalar which represents the probability that $\mathbf{x}$ comes from the real-world data rather than generated data.

Once the parameters of the generative components in VAE/GAN are determined, the generative data will follow the same distribution, which meets the requirement of the perfect sampler. As a result, they can be adopted to design black-box sampling based stegosystem. The pipeline of the stegosystem building on VAE or GAN is presented in Fig. \ref{blackframework}. 

\begin{figure}[h]\centering
	\subfigure[1 bit]{\includegraphics[width=1.3in, angle=0]{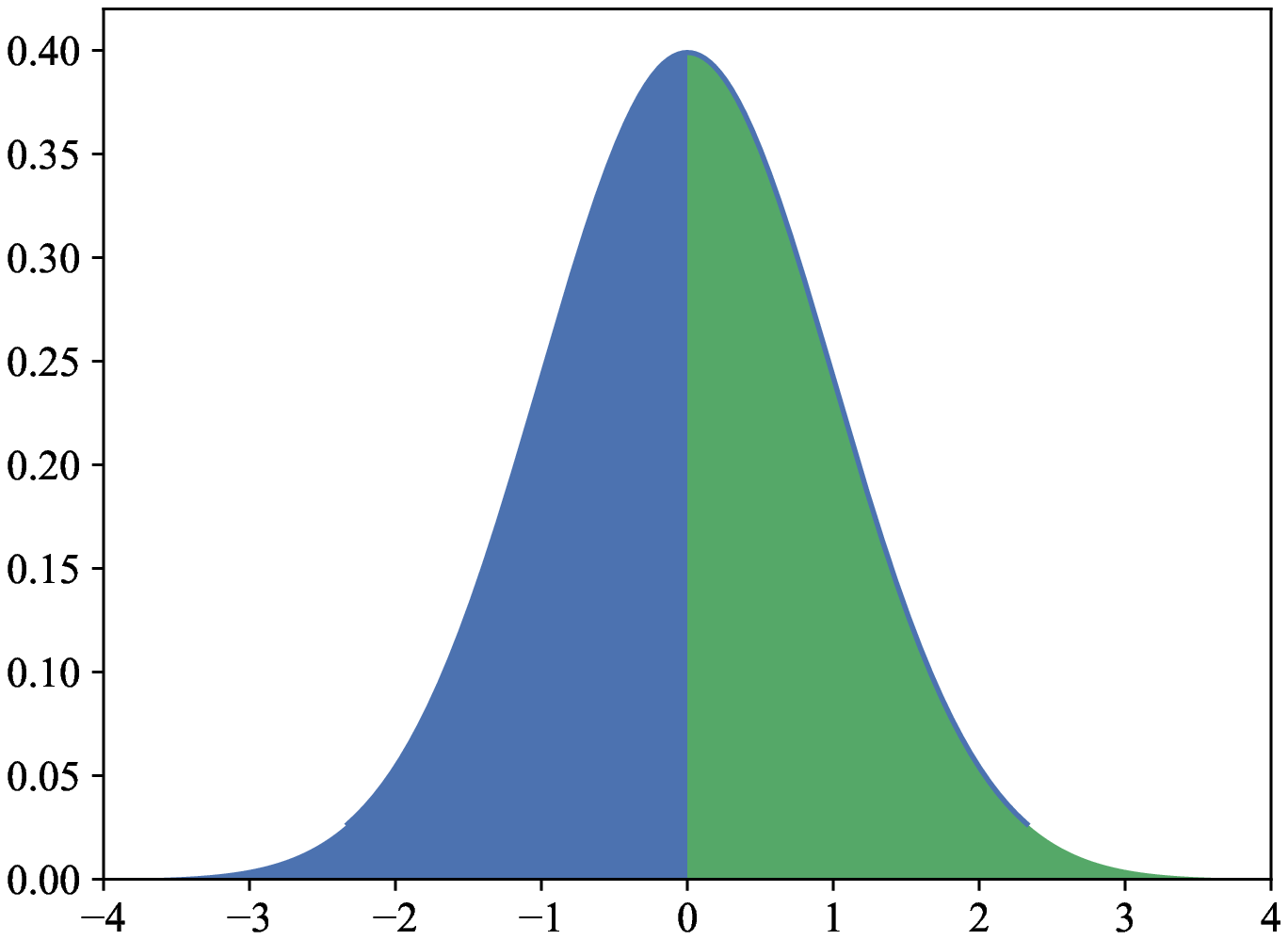}}
	\subfigure[2 bit]{\includegraphics[width=1.3in, angle=0]{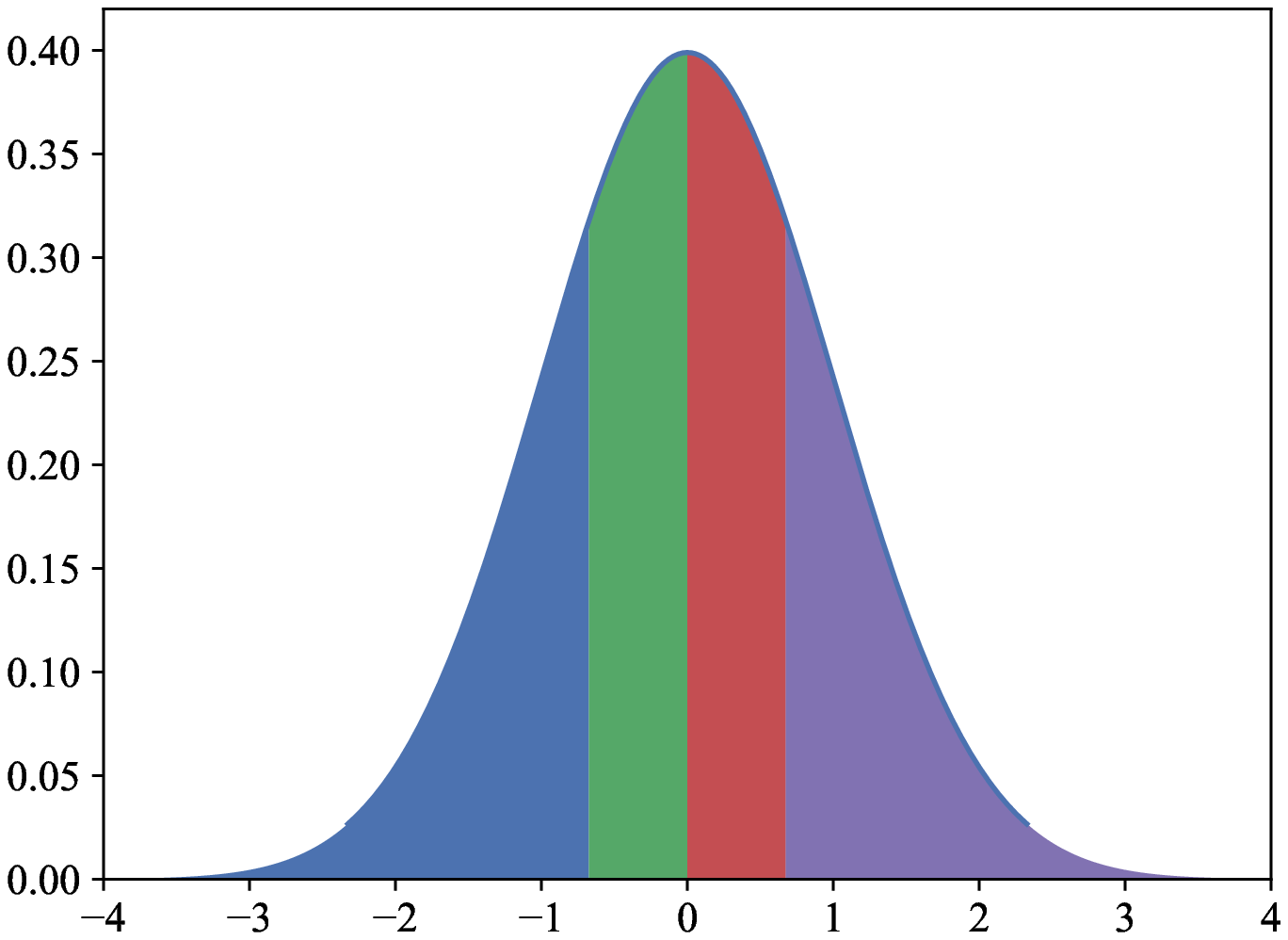}}\\
	\subfigure[3 bit]{\includegraphics[width=1.3in, angle=0]{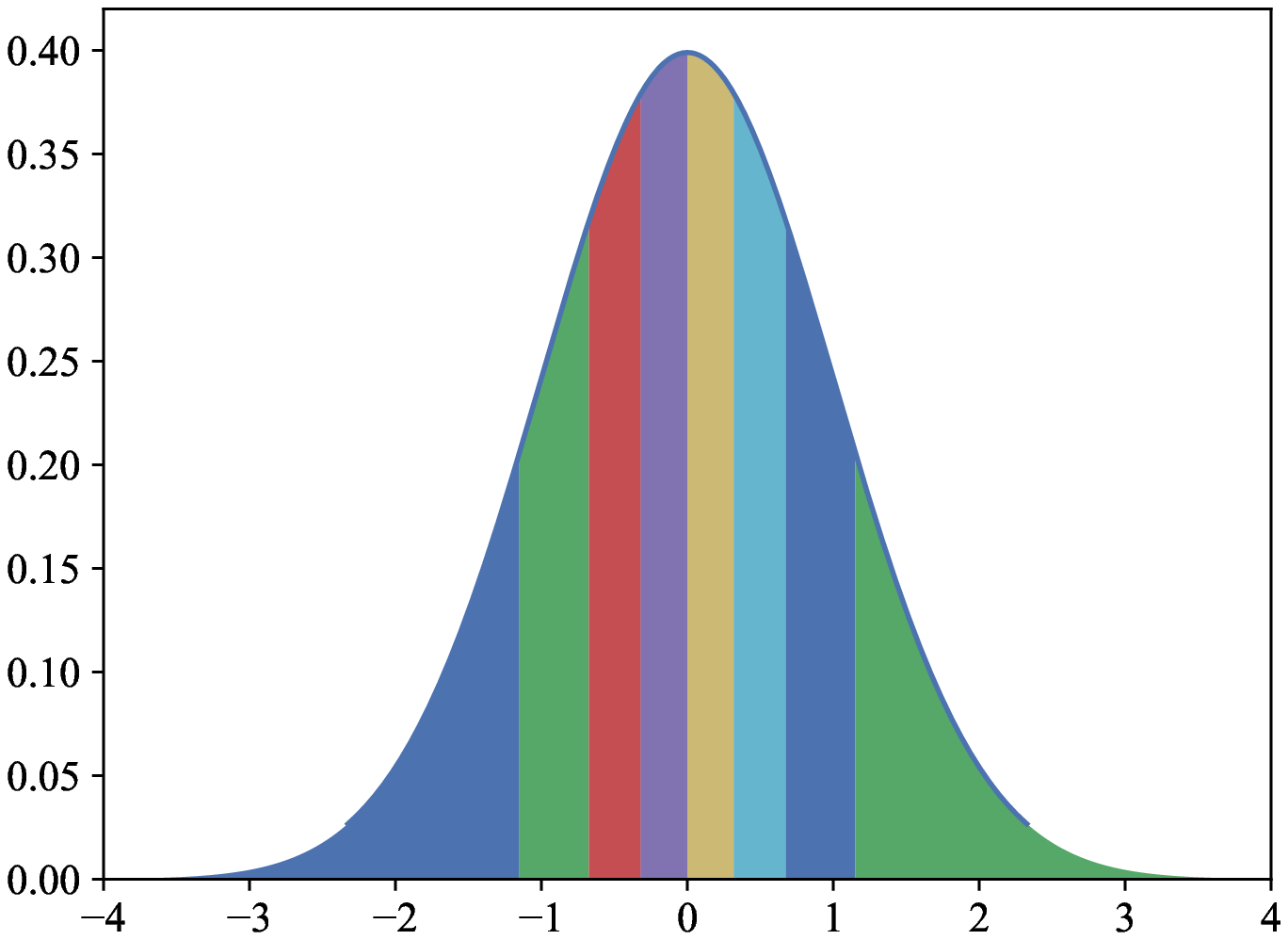}}
	\subfigure[4 bit]{\includegraphics[width=1.3in, angle=0]{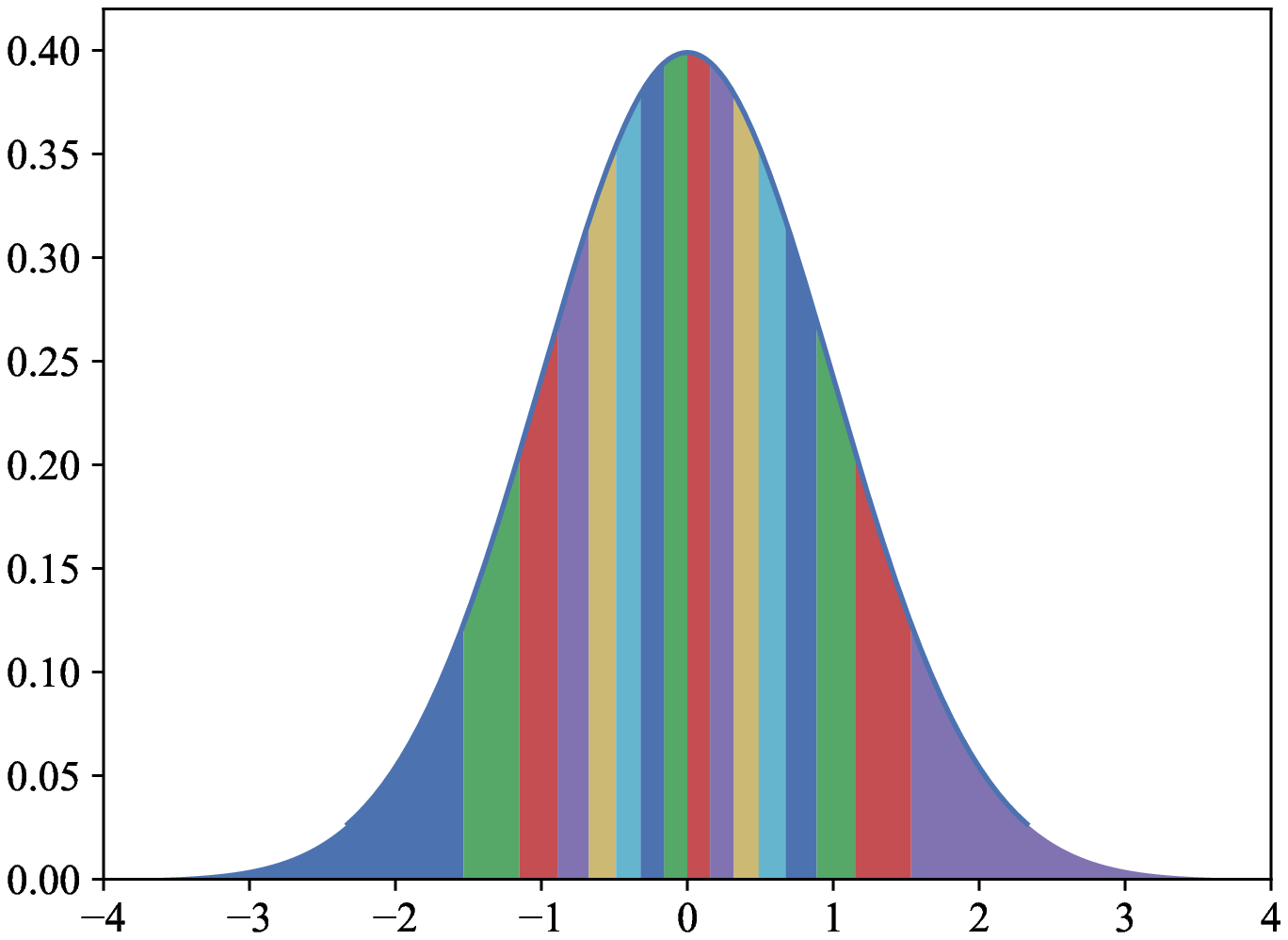}}
	\caption{The examples of division in the module of mapping.}\label{mapping}                
\end{figure}

First, we map the message into latent vectors, and then feed the latent vector $\mathbf{z}$ to a pretrained decoder/generator, which will yield generated \emph{stego} $\mathbf{y}$. Generally, the message is encrypted first, so the encrypted message follows uniform distribution. In VAE or GAN, the latent vectors are constrained to follow normal distribution. Then the mapping module is used to map uniform distribution to normal distribution. Here, we define payload $p$ as the information that each dimension of latent vector carrying. Given $p$ bits of message, we can map it into a random noise using mapping function $\mathcal{M}(m,p)$:
\begin{equation}
z=\mathcal{M}(m,p)=RS\left( \emph{Norm.ppf}\left( \frac{m}{2^{p-1}} \right) ,\emph{Norm.ppf}\left( \frac{m+1}{2^{p-1}} \right)\right),
\end{equation}
where the  $\emph{RS}(x,y)$ is a reject sampling function that will keep randomly sampling a value $z$ from $(-\infty,\infty)$ until $z$ dropping into the interval $(x,y)$. $m$ is the secret message in $p$-ary form transferred from $p$ bits binary message, and $\emph{Norm.ppf}(\cdot)$ is the percent point function (inverse of Cumulative Distribution Function (CDF)) of normal distribution, which can be used to keep the CDF is average divided. Fig. \ref{mapping} gives the example of the interval division of $p=1,2,3,4$, respectively. The bigger the interval is, the easier extracting message is. After the division is determined, reject sampling is utilized to mapping the binary stream into the corresponding interval. Details of the mapping process are given in Algorithm \ref{mappingal}. The module of mapping reconstructed latent vectors $\mathbf{z}'$ to message $\mathbf{m}'$ is the reverse process of Algorithm \ref{mappingal}.

An extra extractor $E$ will be trained to extract the message from the stego, denoted by $\mathbf{z}' = E(\mathbf{y})$ and the extractor $E$ will be sent to receiver before covert communication. The structure of extractor can be similar to the encoder/discriminator with a slight adjustment in the output layer, e.g. setting the dimension of output of the fully connected layer as the dimension of latent vector. The objective function $\mathcal{L}_E$ of the extractor can be set as the mean square error between the original latent vector $\mathbf{z}$ and the reconstructed latent vector $\mathbf{z}'$:
\begin{equation}
\mathcal{L}_E = \lVert \mathbf{z}'-\mathbf{z} \rVert_2.
\end{equation}
Receivers use the extractor network to translate the generated \emph{stego} into reconstructed latent vectors $\mathbf{z}'$ and map $\mathbf{z}'$ into reconstructed message $\mathbf{m}'$ using Demodulator, the inverse process of Modulator.

\begin{algorithm}\label{rs}
	\caption{Mapping function $\mathcal{M}$}
	\label{mappingal}
	\begin{algorithmic}[1]
		\REQUIRE  Payload $p$, message $\mathbf{m}$ of length $n$ 
		\ENSURE latent vectors $\mathbf{v}$
		\STATE Divide the interval (-$\infty$,$\infty$) into $2^p$ subintervals according to the CDF of Uniform distribution.
		\STATE Make every $p$ bits message as one group and compose $\frac{n}{p}$ message groups $\mathbf{G}$.
		\FOR{ each item $i$ in $\mathbf{G}$}
		\REPEAT
		\STATE  Sample $s$ from Uniform distribution.
		\UNTIL {$s$ drops into the corresponding interval of item $i$}.
		\STATE Append $s$ to latent vectors $\mathbf{z}$
		\ENDFOR
	\end{algorithmic}
\end{algorithm}
Since the GAN and VAE share the similar framework, in implementation, we only take Deep Feature Consistent Variational Autoencoder (DFC-VAE)\cite{hou2017deep} as a representative. The structure and the initiation of Extractor are set the same as the encoder in DFC-VAE, the training of Extractor is independent of Encoder and Decoder. In other words, the VAE can be pretrained to generate images. 

\subsection{Provably Secure Stegosystem Using Flow-based Generative Models}\label{s31}
There exists a type of generative model owning the property of reversibility, which is also called flow-based generative model, such as i-ResNet\cite{jacobsen2018revnet}, Glow\cite{kingma2018glow}. In most flow-based generative models, the generative process is defined as:
\begin{equation} 
\vect{z} \sim p_{\boldsymbol{\theta}}(\vect{z}),
\end{equation}
\begin{equation} 
\vect{x}=\vect{g}_{\boldsymbol{\theta}}(\vect{z}),
\end{equation}
where $\mathbf{z}$ is the latent variables and $p_{\boldsymbol{\theta}}$ has a tractable density, such as a spherical multivariate Gaussian distribution: $p_{\boldsymbol{\theta}}(\mathbf{z})=\mathcal{N}(\mathbf{z};0,\mathbf{I})$. $\vect{g}_{\boldsymbol{\theta}}$ is invertible, such that given a datapoint $\vect{x}$, latent-variable inference is done by $\mathbf{z}=\mathbf{f}_{\boldsymbol{\theta}}(\mathbf{x})=\mathbf{g}_{\boldsymbol{\theta}}^{-1}(\mathbf{x})$. The flow-based generative models have achieved considerable visual quality in field of generating images. Since flow-based generative models are reversible, it is very convenient to convert them into steganographic frameworks. The pipeline of stegosystem using flow-based generative model can be expressed as:
\begin{equation} 
\mathbf{m} \xrightarrow{\mathcal{M}} \mathbf{z} \xrightarrow{\mathbf{g}_{\boldsymbol{\theta}}} \mathbf{x} \xrightarrow{\mathbf{g}_{\boldsymbol{\theta}}^{-1}} \mathbf{z}' \xrightarrow{\mathcal{M}^{-1}} \mathbf{m}'.
\end{equation}
The message $\mathbf{m}$ is first mapped into latent vector $\mathbf{z}$, then $\mathbf{z}$ is fed into generative function $\vect{g}_{\boldsymbol{\theta}}$ to obtain the generated \emph{stego} $\mathbf{x}$. The receiver reconstructs the latent variable $\mathbf{z}'$ using $\mathbf{g}_{\boldsymbol{\theta}}^{-1}$, and uses the reverse process of mapping to reconstruct message $\mathbf{m}'$.

Reversibility means the reconstructed latent vector $\mathbf{z}'$ is same as the original latent vector $\mathbf{z}$. Always the mapping function $\mathcal{M}$ is invertible as well. Consequently, the message can be losslessly extracted if the communication channel is noise-free. Unlike the aforementioned framework, this work does not require an extra extractor for it is reversible, which brings great convenience to us. Furthermore, since the process is reversible,  the payload will not affect the accuracy of message extraction. Consequently, the capacity of this framework can reach as big as the file size of image. In our default implementation, Glow is selected as the generative model.

\subsection{Discussion of the Security}\label{sec}
After the training phase, the DFC-VAE and Glow can generate images from latent vector with normal distribution. For innocent users, a random latent vector following normal distribution will be fed to the generative component for generating images. Since the generative component are kept unchanged after training, if the steganographer maps the message into latent vector with normal distribution, the distribution of the generated images will be the same as that generated by innocent users. Consequently, the perfect security can be achieved. It is worth mentioning that the steganographer pursues the distribution of their \emph{stego} images are identical to that of the ordinary generated images rather than natural images.

\section{Provably Secure Steganography on Generative Models With Explicit Probability Distribution}
The compression based stegosystem can be seen as generative steganography, which decompresses the message into \emph{stego}, and can be seen as the dual problem of source coding. Given a perfect compressor, the media can be compressed into bits with the max entropy (following uniform distribution in other words.). The inverse process is decompress uniform bits into media, which coincidentally is similar to the message embedding process, since the message is always encrypted first. Then the perfect security depends on two aspects: perfect compressor and knowing the distribution of the media. There exist many kinds of source coding, such as Huffman codes, Lempel-Ziv codes and Arithmetic coding, asymptotically achieving the theoretical bound. As a result, the tricky problem is that the distribution of natural signal is hard to capture. Luckily, the auto-regressive models with explicit density probability distribution can overcome the problem. 

Autoregressive generative models over high-dimensional data $\mathbf{x}=\left(x_{1}, \dots, x_{n}\right)$ factor the joint distribution as a product of conditionals:
\begin{equation}
p(\mathbf{x})=p\left(x_{1}, \ldots, x_{n}\right)=\prod_{i=1}^{n} p\left(x_{i} | x_{1}, \ldots, x_{i-1}\right).
\label{wavenet1}
\end{equation}
Optionally, the model can condition on additional global information $\vect{h}$ (audio semantic), in which case the conditionals become $p\left(x_{i} | x_{1 : i-1}, \mathbf{h}\right)$.
With the explicit probability distribution of the \emph{cover} object, the source coding can be combined into the generative process for decompressing the message into generated data. Given generated data, if the receivers can repeat the generated process to obtain the probability distribution, they can compress the generated data into message.

\subsection{Provably Secure Stegosystem Using WaveNet}
In the previous work \cite{yang2018provably}, provably secure steganography on image synthesis using autogressive models was presented and shown impressive performance. To extend this work, we design provably secure steganography scheme on audio synthesis system with \emph{WaveNet}. Audio synthesis has natural advantage in this field. The quality of synthesis speech is close to human speech\cite{oord2016wavenet}, and can be easily recognized by speech-to-text system. Furthermore, the system is based on the conditional autogressive models. With different input words, the system can generate the corresponding audio, which means the diversity and the certainty of this system are better than those of the previous work \cite{yang2018provably}.

\begin{figure}[t]
	\centering
	\includegraphics[width=3.5in]{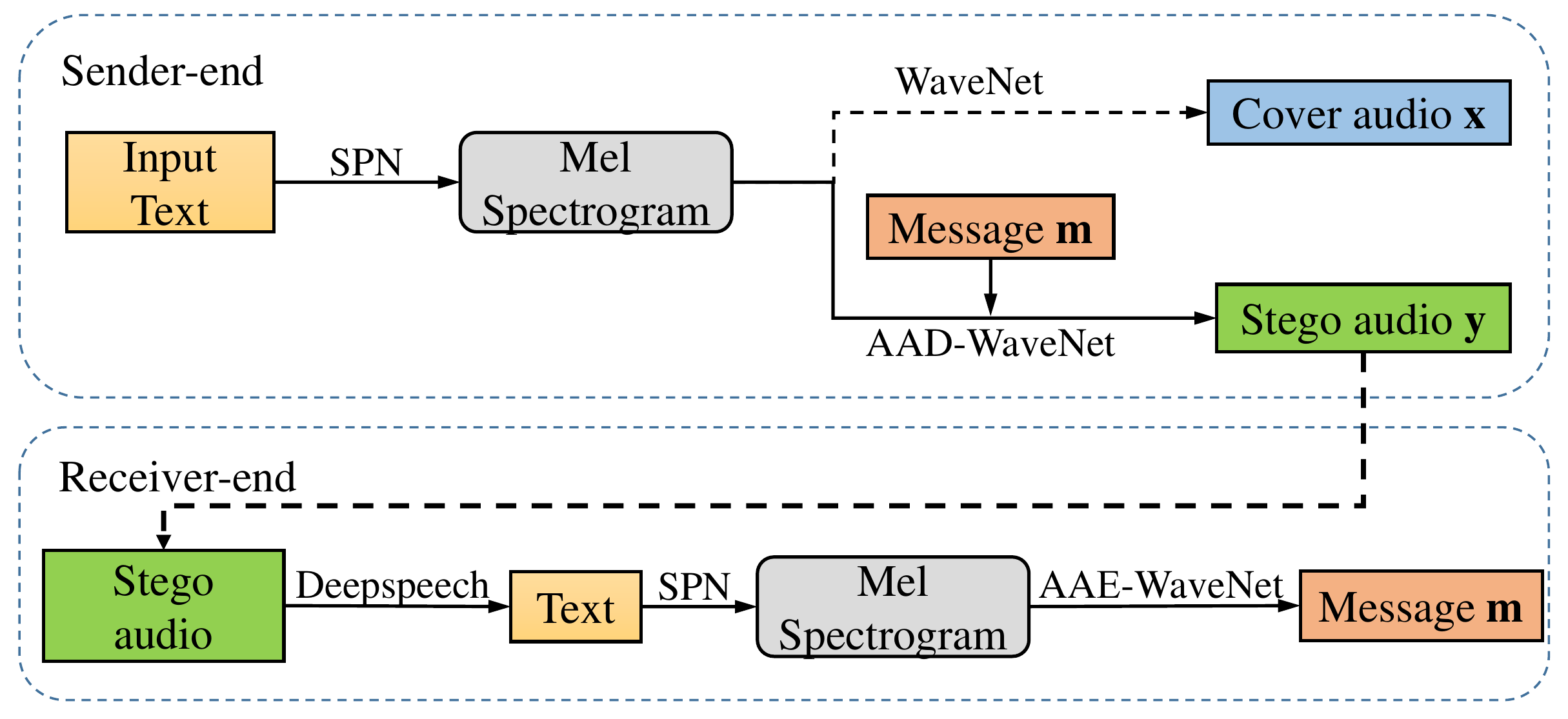}
	\caption{The diagram of generative steganography using \emph{WaveNet} and Arithmetic Coding.} 
	\label{wavdiagram}
\end{figure}
The whole diagram of the stegosystem is shown in Fig. \ref{wavdiagram}. At the sender-end, given the input text, the mel-spectrogram can be predicted by the Spectrogram Prediction Network (SPN). As for ordinary waveform generation, the sample is randomly chosen according to the distribution $p(x_t)$, where the distribution is yielded by the \emph{WaveNet} vocoder. In detail, the probability of the second sample $p\left(x_2\mid x_1, \vect{h}\right)$ is produced by the WaveNet with the mel-spectrogram. As for ordinary waveform generation, the value of the second sample $x_2$ is randomly chosen according to $p\left(x_2\mid x_1, \vect{h}\right)$. Given $x_1, x_2$, the probability distribution of $x_3$ is predicted by the network as $p\left(x_3 \mid x_1, x_2,\vect{h}\right)$, similar process will be repeated until all samples are generated. 

When it comes to steganography, the message embedding process is integrated with the waveform generation in WaveNet vocoder. Since the generation is a sequential generation, adaptive arithmetic decoding (AAD) is utilized to embed message. Given $p(y_2\mid y_1, \vect{h})$ (same as $p(x_2\mid x_1,\vect{h})$), and message $\vect{m}$, a part of message can be embedded and obtained $y_1$ using AAD. 
Then the probability will be updated as $p\left(y_3 \mid y_2,y_1,\vect{h}\right)$. Analogously, the process will be repeated until all samples are generated. The whole process of message embedding is
denoted as: 
\begin{equation}
\vect{y} = AAD\left(p\left(\vect{y} \mid \vect{h}\right),\vect{m}\right).
\end{equation}
where $p\left(\vect{y} \mid \vect{h}\right)$ represents the probability distribution of $\vect{y}$. The process is also named AAD-WaveNet, labeled in Fig. \ref{wavdiagram}. To point out, the message length should be shorter than the entropy of $p\left(\vect{y} \mid \vect{h}\right)$. For simplicity, the message length $L$ will be set as different multiples of a constant value, so that $L$ is not required to send to the receiver. If the real message does not reach the length, just padding 0 to message before encryption. Before the covert communication, the sender should send SPN and WaveNet vocoder to the receiver, or use public models and just tell the receiver where to download. 

The receiver utilizes speech-to-text tool such as DeepSpeech\cite{hannun2014deep}, to recognize speech into text. It is worth mentioning that, the recognition is not perfect, so the sender must verify the generated audio can be correctly recognized before communication. Since owning the same model, $\vect{y}$ and message length $L$, the probability distribution $p(\vect{y})$ of every step will be identical to that generated in the sender-end, so the message can be extracted using adaptive arithmetic encoding (AAE), denoted by:
\begin{equation}
\vect{m} = AAE\left(p\left(\vect{y} \mid \vect{h}\right),\vect{y}\right).
\end{equation}
The details of message embedding and extraction using \emph{AAD} and \emph{AAE} will be elaborated below.

\subsection{Message Embedding and Extraction}

Given the distribution $P_{\text{c}}$ of generative media, the process of embedding message corresponds to source decoding, and extraction corresponds to source encoding. $\mathcal{A}=\{a_1,a_2,...,a_m\}$ is the alphabet for a generative \emph{cover} in a certain order with the probability $\mathcal{P}=\{P(a_1),P(a_2),...,P(a_m)\}$. The cumulative probability of a symbol  can be defined as
\begin{equation}
F\left( a_i\right)=\sum_{k=1}^i P\left( a_k \right).
\end{equation}
Owning these notations, we start to introduce the process of message embedding and extraction.

\textbf{1) Message embedding:} Here,  adaptive arithmetic decoding (AAD) is selected as the source coding. Given the encrypted message $\vect{m}=[m_1m_2m_3...m_L]$, it can be interpreted as a fraction $q$ in the range $[0,1)$ by prepending ``0.'' to it:
\begin{equation}
m_1m_2m_3...m_L  \rightarrow q = 0.m_1m_2m_3...m_L=\sum_{i=1}^L m_i \cdot 2^{-i}.
\label{fraction}
\end{equation}
Following the adaptive arithmetic decoding algorithm, we start from the interval $[0,1)$ and subdivide it into the subinterval $[l, l + h)$ according to the probability $\mathcal{P}$ of the symbols in $\mathcal{A}$, and then append the symbol $a_j$ corresponding to the subinterval in which the dyadic fraction $q$ lies into the \emph{stego} $\vect{y}$:
\begin{equation}
\vect{y} = \vect{y} :: a_j,
\end{equation}
where $::$ represents appending the subsequent symbol into the previous vector. Regularly, the probability $\mathcal{P}$ of symbols will be updated. 
Then calculate the subinterval $[l, h)$ according to the updated probability $\mathcal{P}$ by 
\begin{equation}
h_k = l_{k-1}+(h_{k-1}-l_{k-1})*F(a_{j}),
\end{equation}
\begin{equation}
l_k = l_{k-1}+(h_{k-1}-l_{k-1})*F(a_{j-1}),
\end{equation}
where $h_k$ and $l_k$ are the bound of subinterval in the $k$-th step.
Repeat the process until the fraction $q$ satisfies the constraint:
\begin{equation}
\left\{ \begin{array}{l}
q+\left( 0.5 \right)^{L} \notin  [l_{k},h_{k})\\
q-\left( 0.5 \right)^{L} \notin  [l_{k},h_{k})\\
\end{array} \right. 
\label{contraint}
\end{equation}
This constraint guarantees that the dyadic fraction $q$ is the unique fraction of length $L$ in the interval $[l_{k},h_{k})$, such that the message can be extracted correctly. 
The message length $L$ and the probabilities $\mathcal{P}$ of symbol are shared with the receiver.

To further clarify the scheme of message embedding using arithmetic decoding, we provide a pseudo-code that describes the implementation of message embedding by adaptive arithmetic decoding in Appendix.

\textbf{2) Message extraction:} 
Correspondingly, the message extraction refers to adaptive arithmetic encoding (AAE). On the receiver-end, the interval $[l,h)$ starts from $[0,1)$, and will be subdivided into subintervals of length proportional to the probabilities of the symbols. if $k$-th element $y_j$ corresponds to the symbol $a_j$, update the subinterval as follows:
\begin{equation}
h_k = l_{k-1}+(h_{k-1}-l_{k-1})*F(a_{j}),
\end{equation}
\begin{equation}
l_k = l_{k-1}+(h_{k-1}-l_{k-1})*F(a_{j-1}),
\end{equation}
Repeat the process until the number of steps reaches the length of $\vect{y}$. Finally, find the fraction $q = \sum_{i=1}^nm_{i}2^{-i}$ satisfying $l_n \le q \le h_n$, where $m_{i} \in \{0,1\}$ is the message bit and $n$ is the length of message. 
Analogously, the pseudo code of message extraction is presented in Appendix. 

\subsection{A Simple Example of Message Embedding and Extraction Using AAD and AAE}\label{secproof}
\begin{figure}[t]
	\centering
	\includegraphics[width=3.5in]{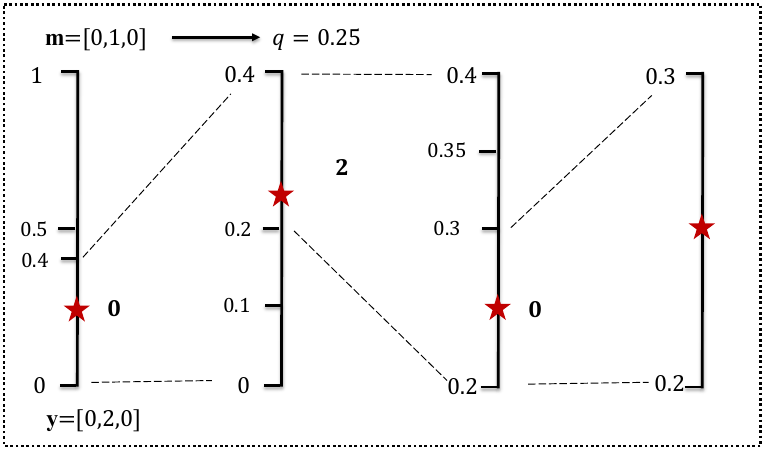}
	\caption{The process of embedding message using adaptive arithmetic decoding (AAD).}
	\label{embed}
\end{figure}

\begin{figure}[t]
	\centering
	\includegraphics[width=3.5in]{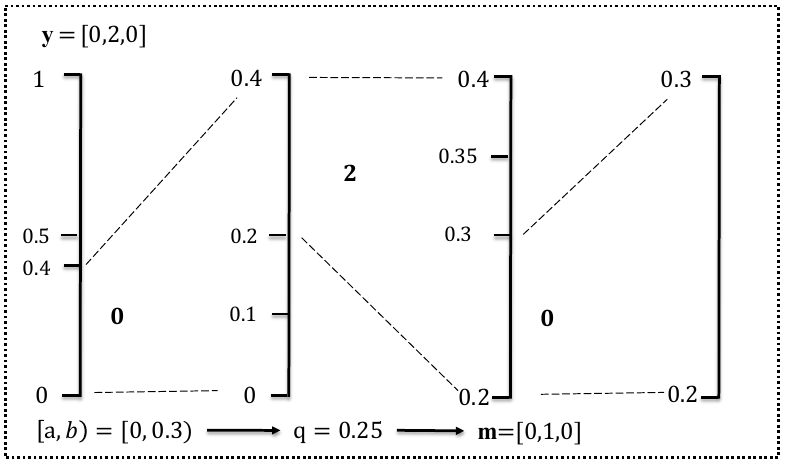}
	\caption{The process of message extraction using adaptive arithmetic encoding (AAE).}
	\label{extract}
\end{figure}
Fig. \ref{embed} presents an example of embedding message using \emph{AAD}.
Given the encrypted message $\mathbf{m} = (m_j)^L$, and the cover distribution $P_{\text{c}}=(p_i)^n$, we will implement embedding by $\mathbf{y}=AAD(\mathbf{m},\mathbf{P}_{\text{c}})$.
Suppose we get $\mathbf{m}=[0,1,0]$ and $P_{\text{c}}=[p_1,p_2,p_3]$ with each $p_j$ as follows:
\begin{equation}
\begin{split}
& p_1 = (0.40,0.10,0.50)\\
& p_2 = (0.25,0.25,0.50)\\
& p_3 = (0.50,0.25,0.25)\\
\end{split}
\label{split}
\end{equation}
Transfer the message $\mathbf{m}=[0,1,0]$ into fraction $q=0.25$. The initial interval is set as [0, 1). Given $p_1 = (0.40,0.10,0.50)$, the three candidate intervals (for $y_i$ to be 0, 1 and 2 respectively) is as follows:
\begin{equation}
I_0=[0,0.4),I_1=[0.4,0.5),I_2=[0.5,1.0).
\end{equation}
As $q \in I_0$, Alice gets $y_1 = 0$ and updates the interval as $[0,0.4)$.
Then given $p_2 = (0.25,0.25,0.50)$, the subintervals can be obtained as:
\begin{equation}
I_0=[0,0.1),I_1=[0.1,0.2),I_2=[0.2,0.4).
\end{equation}
As $q \in I_2$, Alice gets $y_2 = 2$ and updates the interval as $[0.2,0.4)$.
Thereafter, the new subintervals are divided according to $p_3= (0.50,0.25,0.25)$:
\begin{equation}
I_0=[02,0.3),I_1=[0.3,0.35),I_2=[0.35,0.4).
\end{equation}
On the receiver-end, we use adaptive arithmetic encoding $\mathbf{m} = AAE(\mathbf{y},\mathbf{P})$, as shown in Fig. \ref{extract}. The parameters of the generative model is shared with the receiver, so the receiver can obtain the cover's distribution $P_{\text{c}}$. Given $\mathbf{y}$, message length $L$, and the initial interval $[0,1)$, Bob first divide the interval according to $P_1$ in Equation (\ref{split}). Due to $y_1 = 0$, we update the interval as [0,0.4). Given $p_2=(0.25,0.25,0.50)$ and $y_2=2$, we update the interval as [0.2,0.3). Given $p_3=(0.50,0.25,0.25)$ and $y_3=0$. we get the final interval as [0.2,0.3). The final interval contains only a fraction whose binary form is L = 3 bits, denoted by $q=0.25$. So we restore the original message $m = (010)$. 

\subsection{Proof of Asymptotically Perfect Security}\label{secproof}
The secure proof of using Arithmetic coding is discussed in the subsection.
The arithmetic code is prefix free, and by taking the binary representation of $q$ and truncating it to $l\left( c \right) =\lceil \log \frac{1}{P\left( c \right)} \rceil +1$ bits\cite{sayood2017introduction}, we obtain a uniquely decodable code. When it comes to encoding the entire sequence $\vect{c}$, the number of bits $l(\vect{c})$ required to represent $F(\vect{c})$ with enough accuracy such that the code for different values of $\vect{c}$ are distinct is 
\begin{equation}
l\left( \vect{c} \right) =\lceil \log \frac{1}{P\left( \vect{c} \right)} \rceil +1.
\end{equation}
Remember that $l\left( \vect{c} \right)$ is the number of bits required to encode the entire sequence $\vect{c}$. Therefore, the average length of an arithmetic code for a sequence of length $n$ is given by

\begin{equation}
\begin{split}
l_{A^n} &= \sum P(\vect{c})l(\vect{c})  \\
&= \sum P(\vect{c})\left[l\left( c \right) =\lceil \log \frac{1}{P\left( c \right)} \rceil +1\right]\\
&< \sum P(\vect{c})\left[l\left( c \right) =\log \frac{1}{P\left( c \right)} +1+1\right]\\
&= -\sum P(\vect{c})\log{P\left(\vect{c}\right)} +2\sum P(\vect{c})\\
&= H(C^n)+2.
\end{split}
\end{equation}
Given that the average length is always greater than the entropy, the bounds on $l_{A^n}$ are 
\begin{equation}
H(C^n) \le l_{A^n} < H(C^n)+2.
\end{equation}
The length per symbol $l_A$, or rate of the arithmetic code is $\frac{l_A(n)}{n}$. Therefore, the bounds on $l_A$ are 
\begin{equation}
\frac{H(C^n)}{n} \le l_{A} < \frac{H(C^n)}{n}+\frac{2}{n}.
\end{equation}
Also we know that the entropy of the sequence is nothing but the length of the sequence times the average entropy of every symbol\cite{said2004introduction}:
\begin{equation}
H\left(C^{n}\right)=nH(C).
\end{equation}
Therefore,
\begin{equation}
H\left(C\right)\le l_A <H(C) +\frac{2}{n}.
\label{hx}
\end{equation}

In our framework, $P_\text{s}^n$ is the real distribution of $n$ samples generated by the process of message embedding using \emph{AAD}, and $P_\text{c}$ is the target distribution which we are desired to approximate. According to \cite[Theorem 5.4.3]{cover2012elements}, using the wrong distribution $P_\text{s}^n$ for encoding when the true distribution is $P_\text{c}$ incurs a penalty of $D\left( P_{\text{c}}\parallel P_\text{s}^n \right)$. In other words, the increase in expected description length due to the approximate distribution $P_{\text{s}}^n$ rather than the true distribution $P_{\text{c}}$ is the relative entropy $D\left( P_{\text{c}}\parallel P_\text{s}^n \right)$. Directly extended from Eq. (\ref{hx}), $D\left( P_{\text{c}}\parallel P_\text{s}^n \right)$ has upper bound:

\begin{equation}
D\left( P_{\text{c}}\parallel P_\text{s}^n \right)<\frac{2}{n},
\label{dupper}
\end{equation}
and if $n \rightarrow \infty$, then
\begin{equation}
D\left( P_{\text{c}}\parallel P_\text{s}^n \right) \rightarrow 0.
\label{dupper}
\end{equation}

By increasing the length of the sequence, the relative entropy between $P_{\text{c}}$ and $P_{\text{s}}^n$ turns to be 0, meaning that the proposed steganographic scheme can asymptotically achieve perfect security with sufficient elements using arithmetic coding. 


\section{Experiments}
In this section, experimental results and analysis are presented to demonstrate the feasibility and effectiveness of the proposed schemes. 
\subsection{DFC-VAE Stegosystem}
Anime dataset which includes 50,000 color cartoon images is selected as the dataset with cropping and scaling the aligned images to $64 \times 64$ pixels like. We train the encoder and decoder in DFC-VAE\footnote{The source code can be downloaded at \url{https://github.com/svenrdz/DFC-VAE}.} with a batch size of 64 for 150 epochs over the training dataset and use Adam optimizer with an initial learning rate of 0.0005, which is decreased by a factor of 0.5 for the following epochs. Thereafter, map the random message into the latent vector $\mathbf{z}$ and feed $\mathbf{z}$ to the decoder to generate \emph{stego} image $\mathbf{y}$. With the pair latent vectors  and \emph{stego} images, the extractor can be trained with the same setting as encoder. 
\begin{figure}\centering
	\subfigure[cover]{\includegraphics[width=1.5in, angle=0]{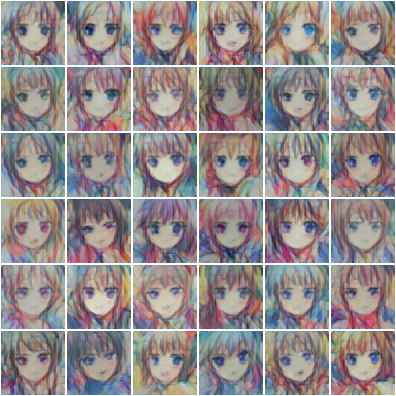}}
	\subfigure[stego]{\includegraphics[width=1.5in, angle=0]{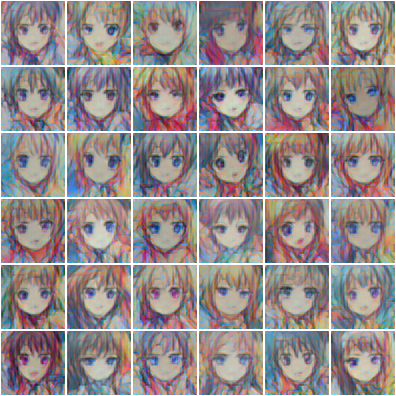}}
	\caption{The \emph{cover} images and \emph{stego} images generated by DFC-VAE stegosystem.}\label{vaesample}                
\end{figure}

\textbf{Quantitative Results and Visualization}
Fig.\ref{vaesample}  shows \emph{cover} samples and \emph{stego} samples obtained by feeding random latent vector and message-driven latent vector to DFC-VAE, respectively. The state-of-the-art image steganalysis methods XuNet\cite{xu2016structural} and SRNet\cite{boroumand2019deep} with a slight modification (modify input channel from 1 to 3 for color images) are used to verify the secure performance of proposed method. Testing error $P_{\text{E}}=\frac{1}{2}(P_{\text{FA}}+P_{\text{MD}})$ is used to quantize the security performance, where $P_{\textrm{FA}}$ and $P_{\textrm{MD}}$ are the false-alarm probability and the missed-detection probability respectively. 10,000 \emph{cover} images and 10,000 \emph{stego} images for different payloads (1,2,3,4 bpp(bit per pixel)) are generated for classifying.

Table \ref{stegadfcvae} shows $P_{\text{E}}$ of DFC-VAE stegosystem, which are all close to 0.5, meaning that the steganalyzer nearly randomly judge that image is \emph{cover} or \emph{stego}. Namely, the DFC-VAE stegosystem is nearly perfect secure. The performance of steganalyzer varies from different datasets, to verify whether the steganalysis methods are powerful, classifying different distribution of this kind of images are added as control experiment. The testing error rates of detecting images generated by unbalance message ($N_{0}:N_{1}=1:2$) are 0.15 and 0.10 by XuNet and SRNet, respectively, which shows validity of steganalyzers.
Accuracy of the secret message recovering has also been explored, and the results are listed in Table \ref{accuracyex}, It can be seen that the larger payload is, the lower accuracy is. large payload means small interval, results in bad fault tolerance.

\begin{table}[]\centering
\caption{The testing error of DFC-VAE stegosystem versus different payloads.}
	\begin{tabular}{c|c|c|c|c}
		\hline
		\hline
		\multirow{2}{*}{Steganalysis} & \multicolumn{4}{c}{Payload (bpp)} \\ \cline{2-5} 
		& 1       & 2      & 3      & 4      \\ \hline
		XuNet\cite{xu2016structural}                         & 0.4952  & 0.4968 & 0.4954 & 0.4944 \\ \hline
		SRNet \cite{boroumand2019deep}                       & 0.4936  & 0.4942 & 0.4986 & 0.4949 \\ \hline
	\end{tabular}

	\label{stegadfcvae}
\end{table}

\subsection{Glow Stegosystem}
Anime dataset also serves as the dataset, while due to the limit computing resource, the images are resized to $32\times32$. The architecture of Glow has a depth of flow $K,$ and number of levels $L$. Here, Glow model was trained with $K=10$, $L=3$, a batch size of 64 for 300 epochs over the training dataset and use Adam optimizer with an initial learning rate of 0.0001. 
\begin{figure}\centering
	\subfigure[cover]{\includegraphics[width=1.5in, angle=0]{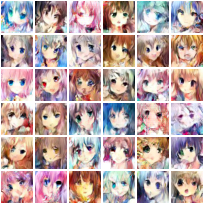}}
	\subfigure[stego]{\includegraphics[width=1.5in, angle=0]{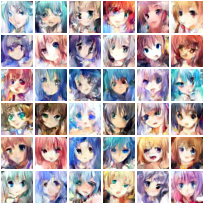}}
	\caption{The \emph{cover} images and \emph{stego} images generated by Glow stegosystem.}\label{glowsample}                
\end{figure}
\begin{table}[]\centering
	\caption{The testing error of Glow stegosystem versus different payloads.}
	\begin{tabular}{c|c|c|c|c}
		\hline
		\hline
		\multirow{2}{*}{Steganalysis} & \multicolumn{4}{c}{Payload (bpp)} \\ \cline{2-5} 
		& 1       & 2      & 3      & 4      \\ \hline
		XuNet\cite{xu2016structural}                         & 0.4968  & 0.4945 & 0.4975 & 0.4986 \\ \hline
		SRNet \cite{boroumand2019deep}                       & 0.4976  & 0.4954 & 0.4964 & 0.4990 \\ \hline
	\end{tabular}

	\label{stegaglow}
\end{table}

\begin{table}[]\centering
	\caption{The message extraction accuracy of DFC-VAE stegosystem and Glow stegosystem.}
	\begin{tabular}{c|c|c|c|c}
		\hline
		\hline
		\multirow{2}{*}{\begin{tabular}[c]{@{}c@{}}Extraction\\ Accuracy\end{tabular}} & \multicolumn{4}{c}{Payload (bpp)}                                                                                    \\ \cline{2-5} 
		& 1          & 2            & 3              & 4               \\ \hline
		DFC-VAE stegosystem      & 97.3\%            & 92.3\%        & 82.87\%           & 66.89\%            \\ \hline
		Glow stegosystem    & \multicolumn{1}{l|}{99.3\%} & \multicolumn{1}{l|}{99.1\%} & \multicolumn{1}{l|}{99.2\%} & \multicolumn{1}{l}{99.2\%} \\ \hline
	\end{tabular}
	
	\label{accuracyex}
\end{table}

\textbf{Quantitative Results and Visualization}
Fig.\ref{glowsample}  shows \emph{cover} samples and \emph{stego} samples obtained by Glow stegosystem. Similar steganalytic experiments are carried out and the results are shown in Table\ref{stegaglow}. The Glow stegosystem are nearly perfect safe, and the accuracy of message recovery approaches 100\%, which show superior to DFC-VAE stegosystem. The image quality is not as good as that in \cite{kingma2018glow}, because the limited computing resources  require us to choose a simple neural network structure. As shown in Table  \ref{accuracyex}, the accuracy rates of the secret message recovering are nearly 100\%, and are independent of the message length, meaning that the capacity of Glow stegosystem is large.

\subsection{WaveNet Stegosystem}
We randomly collect 1,000 short text sentences and transfer them into mel-spectrograms using the SPN\footnote{The architecture of spectrogram prediction network can be downloaded at \url{https://github.com/Rayhane-mamah/Tacotron-2}.} in Tacotron-2\cite{shen2017natural}. 
Then WaveNet vocoder is used for audio waveform generation\footnote{The architecture of WaveNet vocoder can be downloaded at \url{https://github.com/r9y9/wavenet_vocoder}}. The WaveNet vocoder is trained on \emph{CMU ARCTIC} dataset\cite{kominek2004cmu} with 100,000 steps. All the audio clips are stored in the uncompressed WAV format. The audio length ranges from 0.5s to 3s, and the sample rate is 16kHz.

\subsubsection{Steganalysis Features}
The state-of-the-art steganalysis features D-MC\cite{liu2011derivative}, the combined feature of Time-Markov and Mel-Frequency (abbreviated as CTM)\cite{luo2018improved} are selected to evaluate the secure performance. The detectors are trained as binary classifiers implemented using the FLD ensemble with default settings\cite{kodovsky2012ensemble}. A separate classifier is trained for each embedding algorithm and payloads. The ensemble by default minimizes the total classification error probability under equal priors:
\begin{equation}
P_\textrm{E} = \min_{P_{\textrm{FA}}}\frac{1}{2}(P_{\textrm{FA}}+P_{\textrm{MD}}), 
\end{equation}
The ultimate security is qualified by average error rate $\overline{P}_\textrm{E}$ averaged over ten 500/500 database splits, and larger $\overline{P}_\textrm{E}$ means stronger security. 

We also realize other steganographic methods to show the effectiveness of the selected steganalysis features. LSB matching\cite{mielikainen2006lsb} and AACbased\cite{aacstc} algorithms are chosen, where the former is the conventional and the latter is content-adaptive. LSB matching means that if the message bit does not match the LSB of the \emph{cover} element, then one is randomly either added or subtracted from the value of the \emph{cover} pixel element. Otherwise, there is no need for modification.   
AACbased algorithm is simulated at its payload-distortion bound. The distortion of AACbased is defined as the reciprocal of the difference between the original audio and the reconstructed audio through compression and decompression by advanced audio coding. 
\begin{figure}[t]
	\centering
	\includegraphics[width=0.42\textwidth]{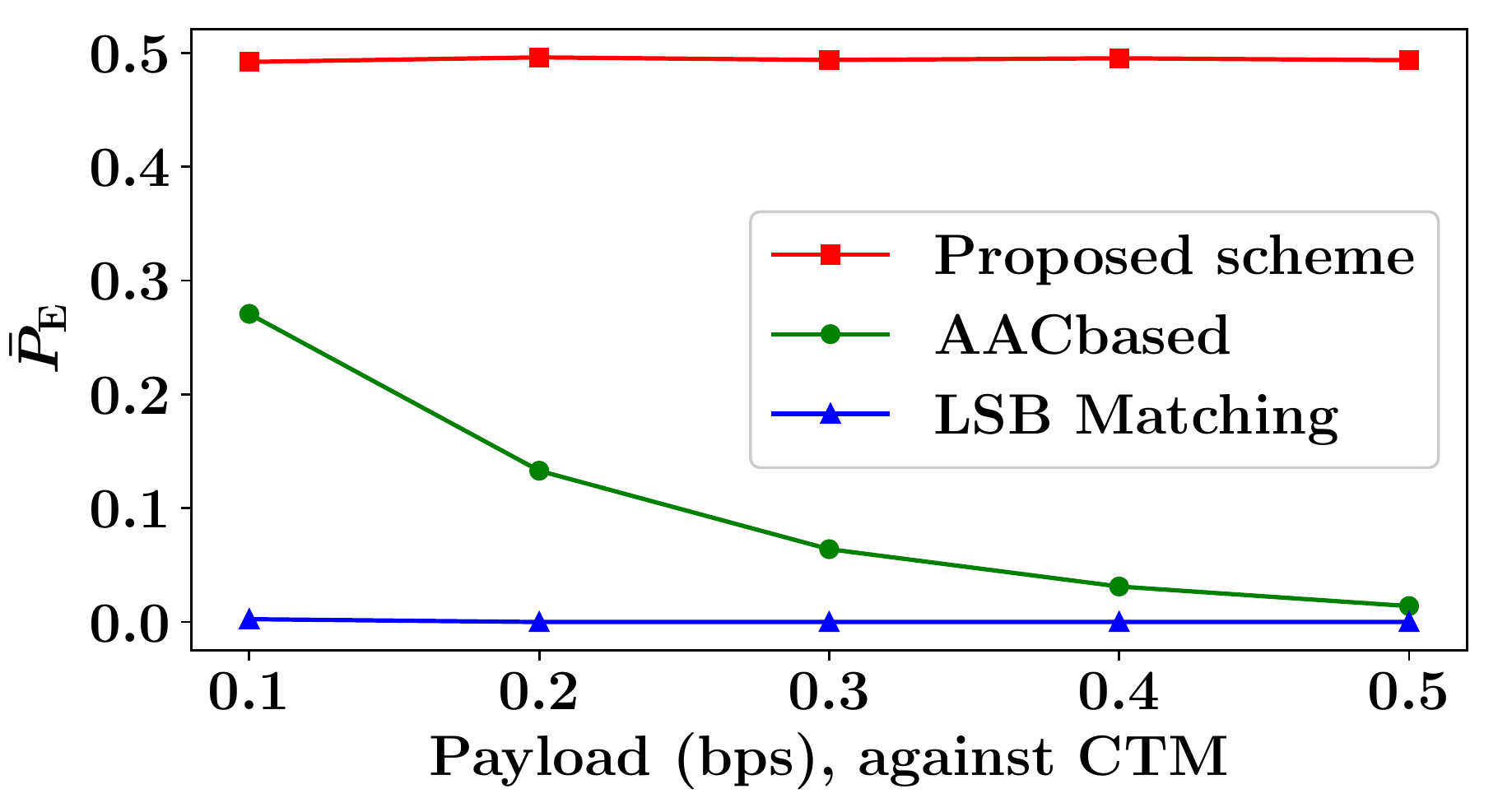}\\
	\caption{\footnotesize{The average detection error rate $\overline{P}_\textrm{E}$ as a function of payload in bits per sample (bps) for steganographic algorithm payloads ranging from 0.1-0.5 bps against CTM.}}
	\label{CTM} 
\end{figure}
\begin{figure}[t]
	\centering
	\includegraphics[width=0.42\textwidth]{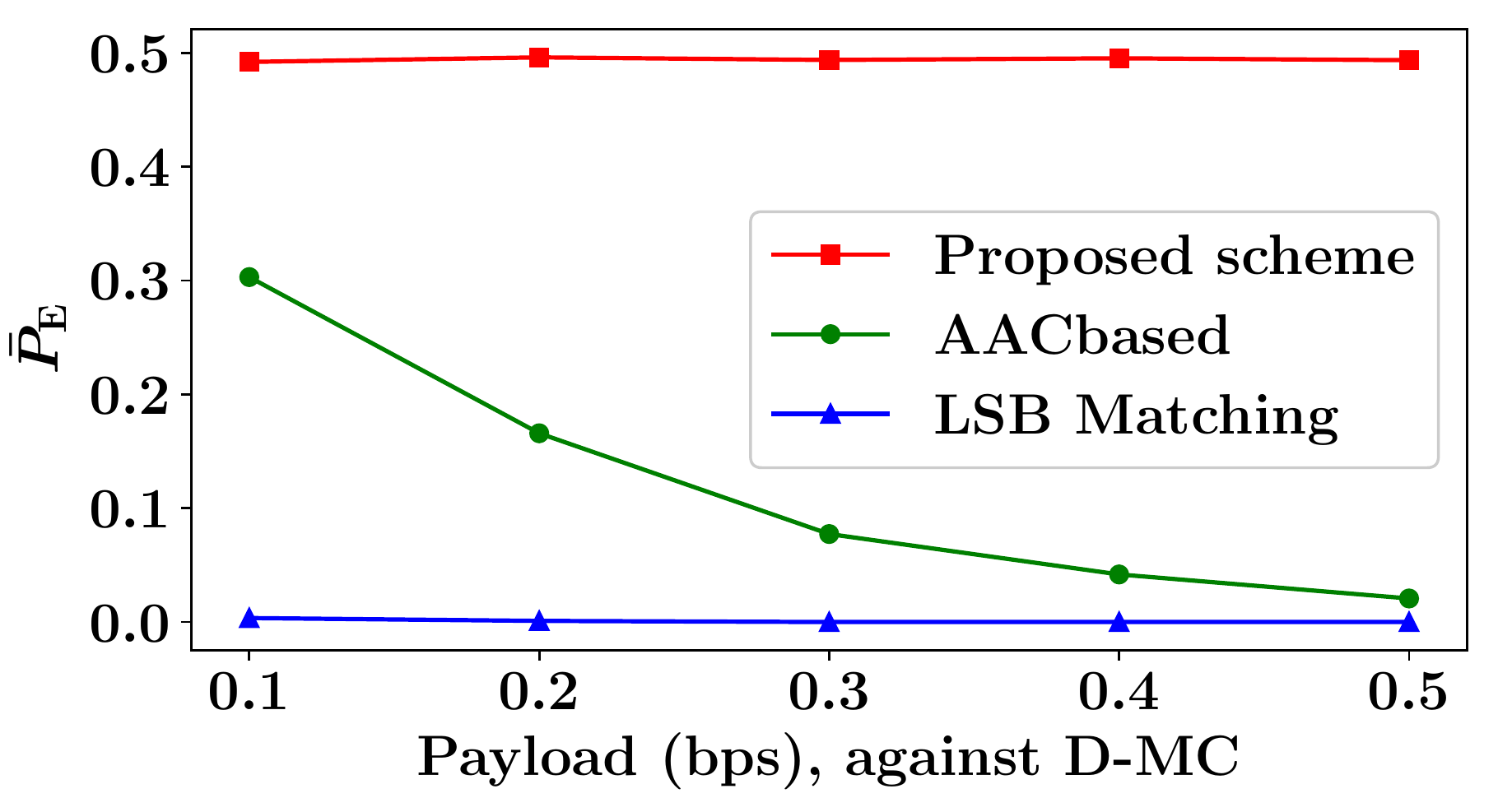}\\
	\caption{\footnotesize{The average detection error rate $\overline{P}_\textrm{E}$ as a function of payload in bits per sample (bps) for steganographic algorithm payloads ranging from 0.1-0.5 bps against D-MC.}}
	\label{DMC} 
\end{figure}

Fig.\ref{CTM} and Fig.\ref{DMC} show the average detection error rate $\overline{P}_\textrm{E}$ as a function of payload in bps for steganographic algorithm payloads ranging from 0.1-0.5 bps against CTM and D-MC. 
It can be observed in Fig. \ref{CTM} and Fig. \ref{DMC} that the $\overline{P}_\textrm{E}$ of AACbased decreases with the increment of payload and turns to be nearly 0\%, and that of LSB matching is always nearly 0\%, showing that the steganalysis is effective with respect to the generated audio. Accordingly, $\overline{P}_\textrm{E}$ of the proposed scheme is nearly 50\%, which means the proposed scheme is nearly perfectly secure. The experimental results verify the security performance as proved in Section \ref{secproof}.

\section{Conclusions}
In this paper, we first review the provably secure steganography, which has two main frameworks: black-box sampling based stegosystem and compression based stegosystem, and conclude the limitation of current work: lack of efficient perfect sampler and cannot model the explicit distribution of natural media. However, the fast development of generative models brings great opportunity to the provably secure steganography. 

Based on generative models without explicit distribution (GAN, VAE, flow-based models), we design block-box sampling based stegosystems. As for GAN/VAE, the generator in GAN and decoder in VAE serve as the perfect sampler, respectively. A message mapping module transfers the message into the latent vector, and then the latent vector will be fed to the perfect sampler to produce the \emph{stego}. An extra extractor network should be trained to extract message. As for flow-based generative model, the generative network can serve as both message embedder and extractor, resulting from its reversible structure. It can be easily found that the flow-based generative models show great convenience in designing stegosystem, for there is no need for an extra extractor. 

For generative models with explicit distribution (autoregressive models), we design compression based stegosystems. Given the distribution of generative media, combining with source coding, we can decompress the message into generative media. The receivers can compress the generative media to extract message.

Take DFC-VAE, Glow, WaveNet as instances, the asymptotically perfectly secure performance is verified by the state-of-the-art steganalysis methods. Additionally, the theoretical proof is given for WaveNet stegosystem using Arithmetic coding.

In our future work, we will explore other effective source encoding schemes and try to transfer them to generative steganographic encoding. Furthermore, other generative media, such as text and video, will be developed under the proposed framework.

\section*{Acknowledgment}
The authors would like to thank Prof. Weiqi Luo from Sun Yat-sen University for providing us the source codes of audio steganalysis. The authors also would like to thank Ryuichi Yamamoto for his valuable advice.

\newpage
\section*{Appendix}
Algorithm \ref{aad} and Algorithm \ref{algaac} give the pseudo codes of message embedding and extraction, respectively.
\begin{algorithm}[H]
	{}\caption{Message embedding using AAD}
	\label{aad}
	\begin{algorithmic}[1]
		\REQUIRE The random message $\vect{m}$, the probability distribution $\mathcal{P}=\{P(a_1),P(a_2),...,P(a_n)\}$, and the cumulative probability $F\left( a_i\right)=\sum_{k=1}^i P\left( a_k \right)$  \ENSURE The stego sequence $\vect{s}$.
		\STATE convert the random message bits $m_1m_2m_3...m_n$ into a fraction $q$ using 
		\begin{equation}
		m_1m_2m_3...m_L  \rightarrow q = 0.m_1m_2m_3...m_L=\sum_{i=1}^L m_i \cdot 2^{-i}.
		\label{fraction}
		\end{equation}
		\STATE $h_0 = 1,l_0 = 0, k = 0$
		\WHILE{$q+0.5^k> h_k\ \&\  q-0.5^k\le l_k $}
		\STATE $k=k+1$
		\STATE subdivide the interval $[l_{k-1}, h_{k-1})$ into subintervals of length proportional to the probability $\mathcal{P}$ of the symbols in cover in the predefined order. The probability $\mathcal{P}$ will be updated when generating next audio sample.
		\STATE take the symbol $a_j$ corresponding to the subinterval in which $q$ lies.
		\STATE $\vect{s} = \vect{s} :: a_j$
		\STATE $h_k = l_{k-1}+(h_{k-1}-l_{k-1})*F(a_{j}).$
		\STATE $l_k = l_{k-1}+(h_{k-1}-l_{k-1})*F(a_{j-1}).$
		\ENDWHILE
		\STATE $output = \vect{s}$
	\end{algorithmic}
\end{algorithm}

\begin{algorithm}[H]
	\caption{Message extraction using AAE}
	\label{algaac}
	\begin{algorithmic}[1]
		\REQUIRE The stego sequence $\vect{s}$, the probability distribution $\mathcal{P}$, the message length $n$ and the CDF $F$.
		\ENSURE The message $\vect{m}$.
		\STATE $h_0$ = 1
		\STATE $l_0$ = 0
		\STATE k = 0
		\WHILE{$h_k \le 2^n$}
		\STATE $k=k+1$
		\STATE subdivide the interval $[l_{k-1}, l_{k-1} + h_{k-1})$ into subintervals of length proportional to the probabilities $\mathcal{P}$ of the symbols in cover (in the predefined order). The probability $\mathcal{P}$ will be updated when generating next audio sample. 
		\STATE $h_k = l_{k-1}+(h_{k-1}-l_{k-1})*F(a_{j}).$
        \STATE $l_k = l_{k-1}+(h_{k-1}-l_{k-1})*F(a_{j-1}).$	
		\ENDWHILE
		\STATE find the fraction $q = \sum_{i=1}^nm_{i}2^{-i}$ satisfying $l_n \le q \le h_n$, where $m_{i} \in \{0,1\}$ is the message bit.
		\STATE $output = \vect{m} = [m_1,m_2,m_3,...m_n]$
		
	\end{algorithmic}
\end{algorithm}

\end{document}